\documentclass{article}
\usepackage{graphicx}

\begin{document}
\title{A systematic study on the binding energy of $\Lambda$ hypernuclei}
\author{LAN Mi-Xiang $^1$\footnote{E-mail:
lanbudai@mail.nankai.edu.cn; lan@mail.ustc.edu.cn},LI Lei
$^2$\footnote{E-mail: lilei@nankai.edu.cn},
NING Ping-Zhi $^2$\footnote{E-mail: ningpz@nankai.edu.cn}\\
1. Interdisciplinary Center of Theoretical Studies,
\\University of Science and Technology of China,\\ Anhui 230026,
P.R.China\\
2. Department of Physics, Nankai University,\\
   Tianjin 300071, P.R.China
\date{}}

\maketitle

\begin{abstract}
We calculated the binding energy per baryon of the $\Lambda$
hypernuclei systematically, using the relativistic mean field
theory(RMF) in a static frame. Some similar properties are found for
most of the $\Lambda$ hypernuclei confirmed by experiments. The data
show that a $\Lambda$ hypernucleus will be more stable, if it is
made by adding a $\Lambda$ hyperon to a stable normal nuclear core,
or by replacing a neutron by a $\Lambda$ hyperon to a stable normal
nuclear core. According to our calculations, existences of some new
$\Lambda$ hypernuclei are predicted under the frame of RMF.
\end{abstract}

The discovery of the strangeness opens a new research field of
nuclear physics with strangeness. The research on the structure and
property of hypernuclei had produced many meaningful results in
1970s. Some methods, for example, the nuclear core model, RMF and
Hartree-Fock-Bogolinbov (HFB), have been used. It has been more than
50 years since the first discovery of a $\Lambda$ hypernuclei by
Polish physicists Marion Danysz and Jerzy Pniewski.$^{[1,2]}$ There
has been great progress in understanding $\Lambda$ hypernuclei since
1970s in experiment and theory. The hypernuclei properties include
the hyperon binding energy $B_\Lambda$$^{[3]}$, the well depth of
$\Lambda$-nuclear potential, hypernuclear spins and life times,
$\Lambda$ hypernuclei spectroscopy$^{[4]}$, the shrinkage of
$\Lambda$ hypernuclear size, $\pi^0$- and $\pi^{+}$-mesonic decays,
excited states of hypernuclei and binding energy per baryon, etc
have been studied. Up to now, 53 hypernuclei have been found,
including 40 single-$\Lambda$ hypernuclei, 3 double-$\Lambda$
hypernuclei, 8 $\Xi^-$ hypernuclei and 2 $\Sigma$ hypernuclei
$^{4}_{\Sigma^{0}}$He, $^{4}_{\Sigma^{+}}$He$^{[5--9]}$. A lot of
experimental data has been accumulated. Systematic studies on the
binding energy of $\Lambda$ hypernuclei are required. Sandulescu
{\it et al.}$^{[10]}$ have used the RMF-BCS approach to treat the
resonant states. They considered the resonant states by the
scattering states located in the vicinity of the resonance energies,
and found that including into the RMF-BCS calculations, only one of
these resonant states gets for the neutron radii, and neutron
separation energies are practically as same as in the more involved
relativistic-Hartree-Bogolinbov (RHB) calculations.

As a convenient method, relativistic mean field(RMF) theory can be
used to study nuclear many-body systems. It has been widely adopted
in predicting the properties of nuclear matter and finite nuclei,
not only normal nuclei but also many kinds of hypernuclei.$^{
[11--13]}$ One of the important aspects in RMF is to determine the
effective meson-baryon couplings. In this work, we choose
NL-SH$^{[14]}$ as the parameters of the meson-baryon coupling
constants, which can well describe the properties of finite nuclei
and hypernuclei.$^{[11--13]}$ Under this frame, binding energies per
baryon of the ground-state $\Lambda$ hypernuclei is calculated with
this uniform set of parameters. After analyzing the theoretical and
experimental data with the RMF, some rules can be summarized.
Considering the characters of the proved hypernuclei (a $\Lambda$
hypernucleus will be more stable, if it is composed of a $\Lambda$
hyperon adding to a stable normal nuclear core, or a $\Lambda$
hyperon replacing a neutron in a stable normal nuclear core) and the
stability of nuclei with magic number and four-periodicity, 71 new
$\Lambda$ hypernuclei were predicted, including 5 new hypernuclei
whose baryon numbers are smaller than 30, the other 66 new
hypernuclei with a baryon numbers larger than 30. The predicted
binding energies imply the possible existence of these new
hypernuclei.

In this letter, the binding energies per baryon(BE/$A$) of the
hypernuclei that have been confirmed by experiments are studied
using the RMF method. The $\Lambda$ hypernuclei with baryon number
$A<30$ and $A>30$ are considered separately. The results of the RMF
are discussed. On basis of the above results, some new hypernuclei
are predicted.

First we study the binding energy per baryon of the hypernuclei
confirmed by experiments$^{[3,4]}$ from Fig.1, we can see some
similar properties with the normal nuclei. When the baryon number is
under 30, the binding energy per baryon present a periodic variety.
The binding energy of the medium hypernuclei is larger than those of
the light and heavy hypernuclei. The binding energy per baryon of
the hypernuclei with the baryon number larger than 120 is around 7.0
MeV and decreases slowly with the increase of the baryon number.


The hypernuclei confirmed by experiments with baryon number under 30
are discussed in the following. For clarity, we plot the binding
energy per baryon of some of the hypernuclei with the baryon number
under 30 in Fig.2. We can see that the hypernuclei with the baryon
number $4n+1$ are bound more tightly than the nearby ones (except
$^9_\Lambda $Be and $^{13}_\Lambda$C, $^{10}_\Lambda$Be is the most
stable hypernuclei among the hypernuclei formed with the normal
nuclear core Be. $^{14}_\Lambda$C is the most stable hypernuclei
among the hypernuclei formed with the normal nuclear core C).
According to this, we can see hyperon is different from nucleon. If
we add a nucleon to a nucleus with the stable structure $4n$, we
will find that the nucleus becomes looser. However, if we add a
hyperon to a nucleus with the stable structure, the situation is
changed. We can see from the curve that most of the hypernuclei with
the baryon number $4n+1$ are bound more tightly than the hypernuclei
with the baryon number $4n$. Considering that $^{20}$Ne, $^{24}$Mg
and $^{28}$Si are the nuclei with stable structure 4n ($\alpha$
cluster nuclear), we calculate the binding energy per baryon of the
$^{20}_{\Lambda}$Ne, $^{21}_{\Lambda}$Ne, $^{24}_{\Lambda}$Mg,
$^{25}_{\Lambda}$Mg, $^{29}_{\Lambda}$Si, as is listed in Table 1.


The results show that the predicted hypernuclei with baryon number
$4n+1$ are bound more tightly than the hypernuclei with baryon
number $4n$. This is in consistent with what we have found for the
hypernuclei confirmed by experiments. BE/$A_{\rm(th)}$ is the
binding energy per baryon calculated by RMF.

We plot the hypernuclei, including both confirmed by experiments and
predicted by RMF, with baryon number under 30 in Fig.3. We can see
that $^{5}_{\Lambda}$He, $^{17}_{\Lambda}$O are bound more tightly
than the nearby hypernuclei. We know that $^{4}$He and $^{16}$O are
not only the $\alpha$ cluster nuclei, but also the nuclei with the
double magic number. Here we notice that a magic number nucleus with
a $\Lambda$ hyperon is a more stable structure.

For the hypernuclei confirmed by experiments with baryon number
larger than 30, we can see from Fig.4 that most of those hypernuclei
(except $^{32}_{\Lambda}$S, $^{33}_{\Lambda}$S, $^{56}_{\Lambda}$Fe,
as is known that S and Fe nuclei are not the magic number nuclei;
however, we know that S and Fe nuclei are also stable; thus
hypernuclei are confirmed by experiments with the S and Fe normal
nuclear core; for the S nucleus, its baryon number is close to 30,
so we can treat it as the $\alpha$ cluster nucleus) are formed
either by adding a $\Lambda$ hyperon to a magic number nuclear core,
or by replacing a neutron by a $\Lambda$ hyperon to a magic number
nucleus. Following this rule, we predicted some new hypernuclei, as
shown in Table 2 and Fig.5


We can see that the variational trend of the binding energy per
baryon of the predicted hypernuclei is consistent with that of the
hypernuclei confirmed by experiments. The binding energy of the
hypernuclei with baryon number between 30 and 65 are nearby 8.8 MeV.
The binding energy per baryon with baryon number between 87 and 93
are around 7.5 MeV. Hypernuclei with the nuclear core Kr seem more
tightly bound, their binding energy per baryon are about 8.7 MeV.
But the binding energy per baryon of the nearby hypernuclei with the
nuclear core Rb are nearby 7.6 MeV. The binding energies per baryon
of the hypernuclei with baryon number larger than 111 are around 7.0
MeV and change slowly with the increase of baryon number.

Finally, we plot all the hypernuclei, including hypernuclei
confirmed by experiments and hypernuclei predicted by RMF, in Fig.6.

In our calculation, the match radius of all the hypernuclei is
unchanged. In fact, it differs from the light hypernuclei to the
heavy hypernuclei. Here we just give out a range of the errors.
BE/$A_{1}$ refers to the binding energy per baryon with the match
radius unchanged. BE/$A_{2}$ refers to the binding energy per baryon
with the match radius being adjusted ($\delta_{x}$ = BE/$A_{1}$ -
BE/$A_{2}$, $\delta_{x}$/x = (BE/$A_{1}$ - BE/$A_{2}$)/(BE/$A_{1}$)
).

We can see from Table 3 that when the baryon number is small (under
41), the error is 0. With the increase of the baryon number, the
error increases. This shows that our parameters are not very
suitable to the medium and heavy hypernuclei. We can see from Table
3, that the variational trend of the binding energy per baryon of
the medium hypernuclei and the heavy hypernuclei with the match
radius being adjusted are more like the variational trend of the
binding energy per baryon of the normal nuclei.



In summary, among those hypernuclei with the baryon number under 30,
$\Lambda$ hypernuclei with baryon number $4n+1$ are bound more
tightly than their nearby ones(except $^{9}_{\Lambda}$Be and
$^{13}_{\Lambda}$C). Most of the $\Lambda$ hypernuclei with the
baryon number larger than 30 are composed with the magic number
nuclei. From the experimental data, we can see that most of the
$\Lambda$ hypernuclei confirmed by experiments are made either by
adding a $\Lambda$ hyperon to a stable normal nuclear core, or by
replacing a neutron by a $\Lambda$ hyperon to a stable normal
nuclear core. Then we have predicted some new $\Lambda$ hypernuclei
and studied their binding energies per baryon. The results of RMF
show that these predicted new $\Lambda$ hypernuclei are possible to
be formed.

\begin{table}
\caption{Predicted hypernuclei with baryon number under 30 and their
binding energy per particle}

\begin{tabular}{p{1.8cm}p{1.1cm}p{1.4cm}p{1.8cm}p{1.1cm}p{1.4cm}}
\hline \small{Predicted Hypernuclei}&\small{Baryon number}&\small{BE/$A_{(th)}$ (MeV)}&\small{Predicted Hypernuclei}&\small{Baryon number}&\small{BE/$A_{(th)}$ (MeV)}\\
\hline$^{20}_\Lambda$Ne&20&7.635&$^{25}_\Lambda$Mg&25&7.950\\
\hline$^{21}_\Lambda$Ne&21&7.714&$^{29}_\Lambda$Si&29&8.598\\
\hline$^{24}_\Lambda$Mg&24&7.723&-&-&-\\
\hline
\end{tabular}
\end{table}

\begin{table}
\caption{Predicted hypernuclei with baryon number larger than 30 and
their binding energy per baryon}
\begin{tabular}{p{1.8cm}p{1.1cm}p{1.4cm}p{1.8cm}p{1.1cm}p{1.4cm}}
\hline \small{Predicted Hypernuclei}&\small{Baryon number}&\small{BE/$A_{(th)}$ (MeV)}&\small{Predicted Hypernuclei}&\small{Baryon number}&\small{BE/$A_{(th)}$ (MeV)}\\
\hline$^{36}_\Lambda$S&36&8.637&$^{112}_\Lambda$Sn&112&6.925\\
\hline$^{37}_\Lambda$S&37&8.704&$^{113}_\Lambda$Sn&113&6.961\\
\hline$^{38}_\Lambda$Ar&38&8.597&$^{114}_\Lambda$Sn&114&6.996\\
\hline$^{39}_\Lambda$Ar&39&8.728&$^{115}_\Lambda$Sn&115&7.032\\
\hline$^{48}_\Lambda$Ca&48&8.855&$^{116}_\Lambda$Sn&116&7.030\\
\hline$^{49}_\Lambda$Ca&49&8.880&$^{117}_\Lambda$Sn&117&7.028\\
\hline$^{50}_\Lambda$Ti&50&8.843&$^{118}_\Lambda$Sn&118&7.027\\
\hline$^{51}_\Lambda$Ti&51&8.904&$^{119}_\Lambda$Sn&119&7.026\\
\hline$^{52}_\Lambda$V&52&8.909&$^{120}_\Lambda$Sn&120&7.026\\
\hline$^{52}_\Lambda$Cr&52&8.816&$^{121}_\Lambda$Sn&121&7.026\\
\hline$^{53}_\Lambda$Cr&53&8.910&$^{122}_\Lambda$Sn&122&7.026\\
\hline$^{54}_\Lambda$Fe&54&8.774&$^{123}_\Lambda$Sn&123&7.027\\
\hline$^{55}_\Lambda$Fe&55&8.897&$^{124}_\Lambda$Sn&124&7.028\\
\hline$^{58}_\Lambda$Ni&58&8.856&$^{125}_\Lambda$Sn&125&7.030\\
\hline$^{59}_\Lambda$Ni&59&8.847&$^{132}_\Lambda$Sn&132&7.066\\
\hline$^{60}_\Lambda$Ni&60&8.841&$^{133}_\Lambda$Sn&133&7.074\\
\hline$^{61}_\Lambda$Ni&61&8.838&$^{136}_\Lambda$Xe&136&6.949\\
\hline$^{62}_\Lambda$Ni&62&8.839&$^{137}_\Lambda$Xe&137&6.968\\
\hline$^{63}_\Lambda$Ni&63&8.844&$^{138}_\Lambda$Ba&138&6.886\\
\hline$^{64}_\Lambda$Ni&64&8.853&$^{139}_\Lambda$Ba&139&6.911\\
\hline$^{65}_\Lambda$Ni&65&8.865&$^{140}_\Lambda$La&140&6.893\\
\hline$^{86}_\Lambda$Kr&86&8.778&$^{140}_\Lambda$Ce&140&6.845\\
\hline$^{87}_\Lambda$Kr&87&8.796&$^{141}_\Lambda$Ce&141&6.874\\
\hline$^{87}_\Lambda$Rb&87&7.680&$^{141}_\Lambda$Pr&141&6.823\\
\hline$^{88}_\Lambda$Rb&88&7.587&$^{142}_\Lambda$Pr&142&6.855\\
\hline$^{88}_\Lambda$Sr&88&7.657&$^{142}_\Lambda$Nd&142&6.800\\
\hline$^{89}_\Lambda$Sr&89&7.564&$^{143}_\Lambda$Nd&143&6.834\\
\hline$^{89}_\Lambda$Y&89&7.601&$^{144}_\Lambda$Sm&144&6.753\\
\hline$^{90}_\Lambda$Y&90&7.510&$^{145}_\Lambda$Sm&145&6.790\\
\hline$^{90}_\Lambda$Zr&90&7.545&$^{209}_\Lambda$Pb&209&6.726\\
\hline$^{91}_\Lambda$Zr&91&7.455&$^{210}_\Lambda$Bi&210&6.711\\
\hline$^{92}_\Lambda$Mo&92&7.378&$^{210}_\Lambda$Po&210&6.689\\
\hline$^{93}_\Lambda$Mo&93&7.291&$^{211}_\Lambda$Po&211&6.696\\
\hline
\end{tabular}
\end{table}

\begin{table}
\caption{\leftline{Errors of our calculation}}
\begin{tabular}{ccccc}
\hline Hypernuclei&BE/$A_{1}$(MeV)&BE/$A_{2}$(MeV)&$\delta_{x}$&$\delta_{x}/x$\\
\hline$^{10}_\Lambda$Be&-6.029&-6.029&0&0\\
\hline$^{12}_\Lambda$B&$-7.184$&$-7.184$&$0$&$0$\\
\hline$^{12}_\Lambda$C&$-6.937$&$-6.937$&$0$&$0$\\
\hline$^{13}_\Lambda$C&$-7.801$&$-7.801$&$0$&$0$\\
\hline$^{28}_\Lambda$Si&$-8.294$&$-8.294$&$0$&$0$\\
\hline$^{41}_\Lambda$Ca&$-8.744$&$-8.744$&$0$&$0$\\
\hline$^{51}_\Lambda$V&$-8.832$&$-8.792$&$-0.04$&$0.0045$\\
\hline$^{89}_\Lambda$Y&$-7.601$&$-8.825$&$1.224$&$-0.1610$\\
\hline$^{139}_\Lambda$La&$-6.866$&$-6.866$&$0$&$0$\\
\hline$^{208}_\Lambda$Pb&$-6.721$&$-5.850$&$-0.871$&$0.1296$\\
\hline
\end{tabular}
\end{table}

\begin{figure}
  \centering
  \includegraphics{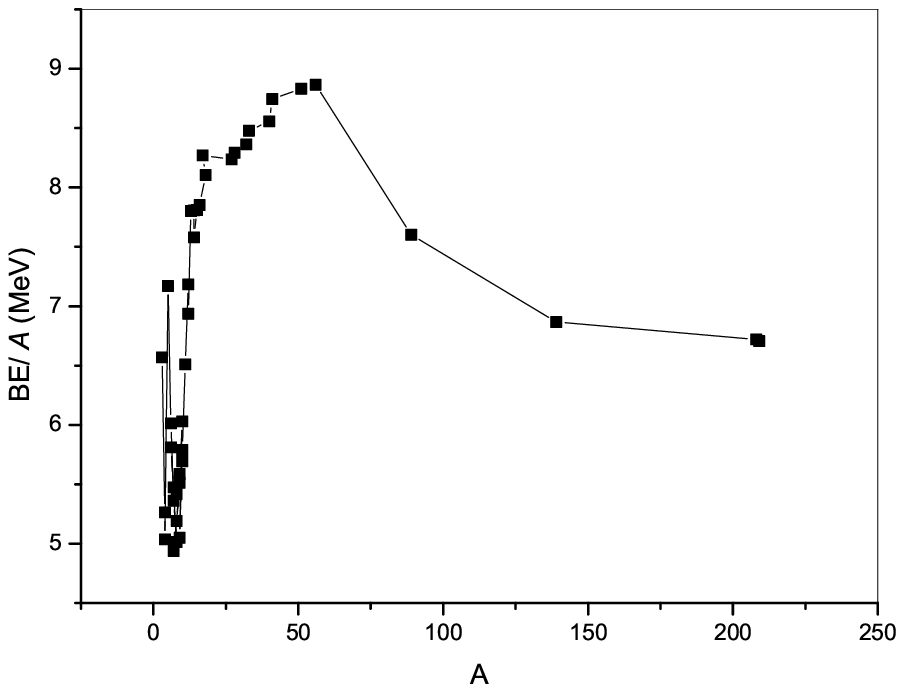}
  \renewcommand{\figurename}{Fig.}
  \caption{\small{Binding energy per baryon of the hypernuclei confirmed by experiments}}
\end{figure}

\begin{figure}
  \centering
  \includegraphics{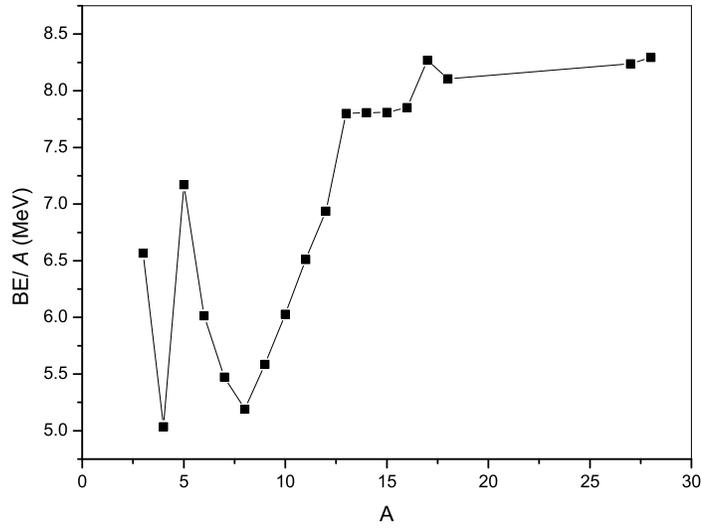}
  \renewcommand{\figurename}{Fig.}
  \caption{\small{Binding energy per baryon of some hypernuclei with
baryon number under 30}}
\end{figure}

\begin{figure}
  \centering
  \includegraphics{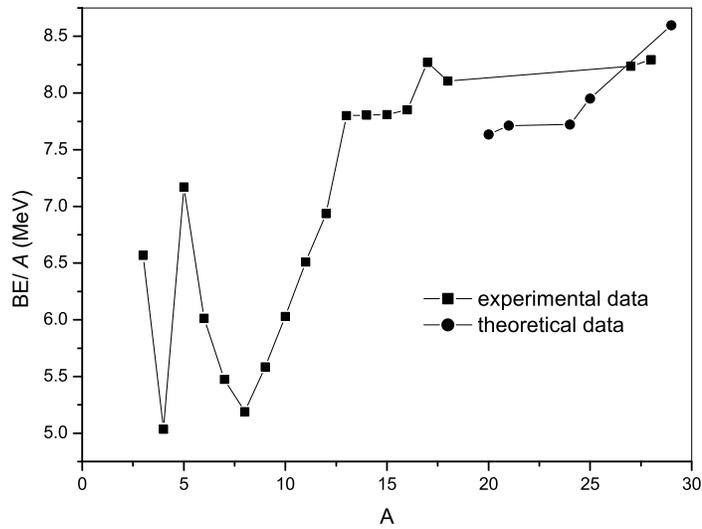}
  \renewcommand{\figurename}{Fig.}
  \caption{\small{Binding energy per baryon of some hypernuclei confirmed by experiments and all the predicted hypernuclei with the baryon
number under 30}}
\end{figure}

\begin{figure}
  \centering
  \includegraphics{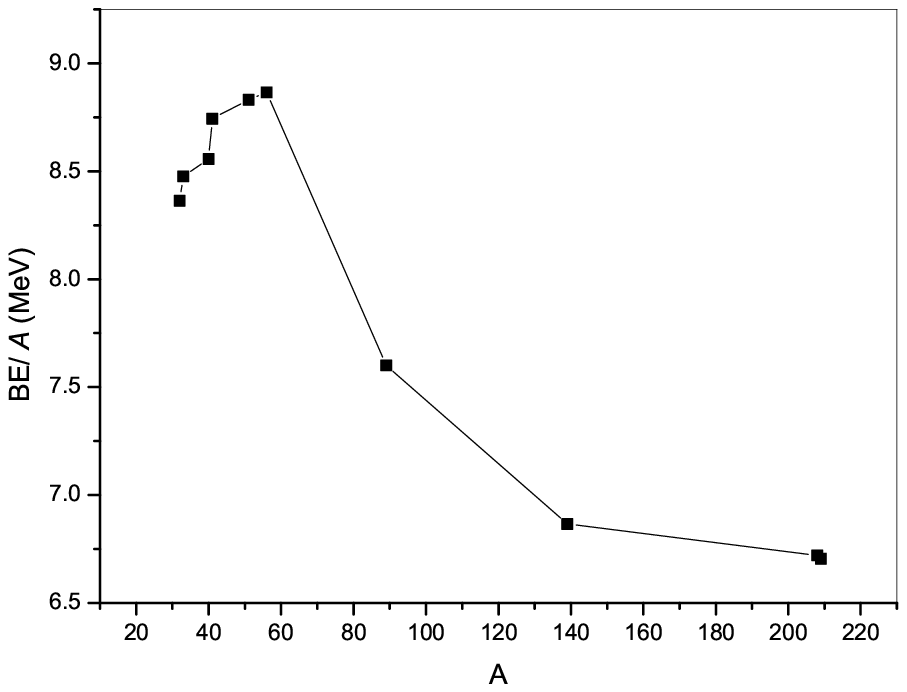}
  \renewcommand{\figurename}{Fig.}
  \caption{\small{Binding energy per baryon of the hypernuclei
confirmed by experiments}}
\end{figure}

\begin{figure}
  \centering
  \includegraphics{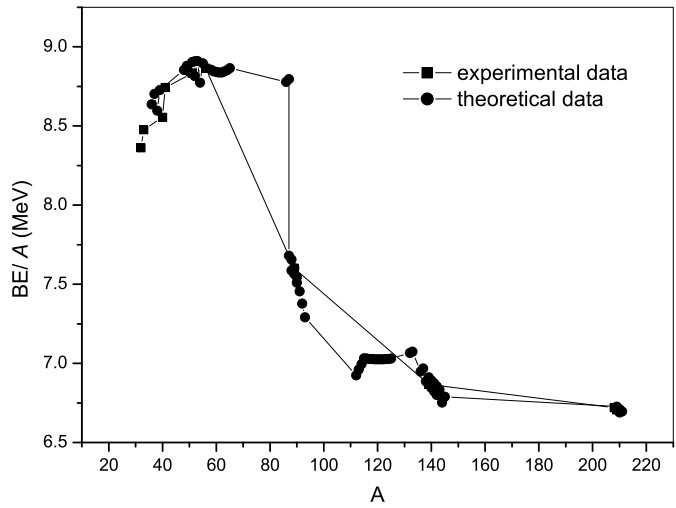}
  \renewcommand{\figurename}{Fig.}
  \caption{\small{Binding energy per baryon of the
hypernuclei with the baryon number larger than 30}}
\end{figure}

\begin{figure}
  \centering
  \includegraphics{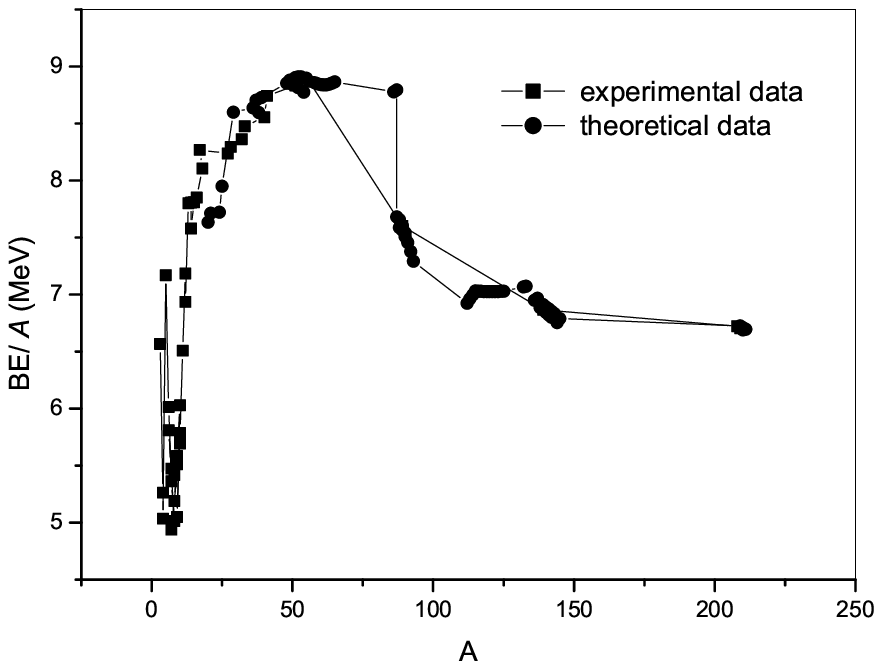}
  \renewcommand{\figurename}{Fig.}
  \caption{\small{Binding energy per baryon of the hypernuclei confirmed by experiments and predicted by RMF}}
\end{figure}


\begin{thebibliography}{99}
\bibitem{DDH} Davis D H 2005 Nucl.Phys.A 754 3
\bibitem{DM} Danysz M and Pniewski J 1953 Phil.Mag. 44 348
\bibitem{SC} Samanta C, Chowdhury P R and Basu D N 2006 Jour.Phys.G:Nucl.Part.Phys. 32 363
\bibitem{HO} Hashimoto O and Tamura H 2006 Prog.Part.Nucl.Phys. 57 564
\bibitem{NT} Nagae T, Miyachi T, Fukuda T \emph{et al} 1998 Phys.Rev.Lett. 80 1605
\bibitem{OutaH} Outa H, Yamazaki T, Iwasaki M and Hayano R S 1994 Prog.Theor.Phys.Suppl. 117 177
\bibitem{OH} Outa H 1996 Hyperfine interactions. 103 227
\bibitem{HRS} Hayano R S 1992 Nucl.Phys.A 547 151c
\bibitem{SRI} Sawafta R I 1998 Nucl.Phys.A 639 103c
\bibitem{SN} Sandulescu N, Geng L S, Toki H, and Hillhouse G C 2003 Phys.Rev.C 68 054323
\bibitem{YHT} Tan Yu-Hong, Zhong Xian-Hui, Cai Chong-Hai and Ning Ping-Zhi 2004 Phys.Rev.C 70 054306
\bibitem{ZhongXH} Zhong X H, Peng G X, Li L and Ning P Z 2006 Phys.Rev.C 74 034321
\bibitem{ZXH} Zhong X H, Tan Y H, Peng G X, Li L and Ning P Z 2005 Phys.Rev.C 71 015206
\bibitem{S} Sharma M M and Nagarajan M A 1993 Phys.Lett.B 312 377
\end{thebibliography}
\end{document}